\documentclass[12pt]{article}
\usepackage{epsfig}
\def\ArtWork#1{\noindent\hfill\epsfbox{#1}\hfill}

\begin{document}

\addtolength{\topmargin}{-2cm}
\addtolength{\oddsidemargin}{-1cm}

\markboth{Arijit Laha}{On building Information Warehouses}

\title{On building Information Warehouses}

\author{ARIJIT LAHA\\ SETLabs\\ Infosys Technologies Ltd.\\ Hyderabad}

%
%
%

\date{}
\maketitle
\begin{abstract}
One of the most important goals of information management (IM) is supporting the knowledge workers in performing their works. In this paper we examine issues of relevance, linkage and provenance of information, as accessed and used by the knowledge workers. These are usually not adequately addressed in most of the IT based solutions for IM. Here we propose a non-conventional approach for building information systems for supporting the knowledge workers which addresses these issues. The approach leads to the ideas of building Information Warehouses (IW) and Knowledge work Support Systems (KwSS). Such systems can open up potential for building innovative applications of significant impact, including those capable of helping organizations in implementing processes for double-loop learning.
\end{abstract}


\section{Introduction}
Effective ``information management", as the means of capture and processing of information as well as accruing maximum benefit out of the available information, has become one of the greatest concerns in modern organizations. However, ``information management" or IM, being an umbrella term, has many facets. One major aspect of IM involves capture and processing of ``structured information" or ``data". Data, representing ``facts" captured or generated in the course of various operational activities, is the life-blood of an organization. Though their main utilization is in enabling the operational activities, they can also form basis of strategic activities, where the data is processed and analyzed by means various sophisticated tools to provide insights to the decision-makers. Computer-based information technology (IT) provides powerful tools and techniques in this area, including DBMS, data warehousing, OLAP, data mining, many other specialized analytic tools and various web-based technologies.

However, the \emph{data} or \emph{structured information} is not the only kind of information available in an organization. The other kind of information, often amounting much greater in volume as well as importance than data, is consisted of \emph{unstructured information}, also called \emph{unstructured data}, typically contained \emph{textual documents} of various kinds. Henceforth, unless otherwise indicated, we shall call it simply \emph{information}. Information is a vital class of resource to the organizations for enabling their \emph{knowledge workers}\cite{drucker} to perform efficiently. Information, in the stated sense, is very closely related to the \emph{knowledge} possessed by individual human beings. According to Alavi and Leidner\cite{alavi},
  \begin{quote}
  ``...\emph{information} is converted to \emph{knowledge} once it is processed \emph{in the mind of individuals} and knowledge becomes information once it is \emph{articulated} and \emph{presented} in the form of text, graphics, words, or other symbolic forms".
\end{quote}
Naturally, effective management of information, for supporting the knowledge workers in their tasks, is a complicated problem, which can not be dealt with information technology alone. This problem has given rise to a very active field of  study, known as \emph{Knowledge Management} or \emph{KM}, an area of multidisciplinary study, straddling major disciplines of information systems, management, organizational learning and strategy \cite{bray}, economics, philosophy and epistemology, computer science and sociology \cite{earl}.

The researchers in disciplines like organizational study and management usually attempt to provide a \emph{holistic view} of KM in organizations. For example, Nonaka\cite{Nonaka} and Markus\cite{Markus} proposed characterization of various activities and processes related to the knowledge-intensive works in organizations. Though, the above works do not explicitly demand use of IT-based tools rather they suggest areas where IT can contribute. The issue of using information systems (IS) for KM was more extensively discussed by Alavi and Leidner\cite{alavi} from the perspective of organizational studies. They discussed the possible roles of IS  with respect to helping the knowledge workers in performing effectively in four categories of knowledge processes \emph{knowledge creation}, \emph{knowledge storage/retrieval}, \emph{knowledge transfer} and \emph{knowledge application}. Unfortunately, in IT/IS community the KM-related issues seldom receive attention from such holistic viewpoint. Despite a glut of IT-based KM tools/solutions in the market, each of them typically cater some narrowly-focused areas of activities. Even in academia, the term ``knowledge" tends to be used rather indiscriminately and thus often obstruct us to address the core issues. An excellent account of such practices can be found in Wilson's article ``The nonsense of knowledge management"\cite{Wilson} and some other articles published in the same issue.

In the following, first we examine the relationships among \emph{information}, \emph{knowledge} and \emph{information systems}. Then we highlight some crucial requirements of the knowledge workers with respect to the access and usage of information. The following section outlines a novel scheme of information re-organization and archival, called ``Information Warehouse" (IW) to facilitate the users' requirement. In the next section we outline a framework called ``Knowledge work Support System" (KwSS) for building comprehensive IS environments based on IW.

\section{Building IS for supporting \emph{knowledge work}}

The  \emph{knowledge-intensive tasks} or \emph{knowledge works} (KW) are defined by Mack et al.\cite{mack} as
\begin{quote}
``...solving problems and accomplishing goals by gathering, organizing, analyzing, creating, and synthesizing information and expertise".
\end{quote}
It can be easily seen that for an organization the crucial issue is to build and maintain the capability of getting required knowledge works performed. To do so, it must engage workers with relevant knowledge and expertise. Further, it needs to provide the workers with environment and resources to carry out their works. KM is the means to achieve them effectively. As noted earlier, the scope of the problem is much diverse and broader than what can be addressed by information systems alone. Nevertheless, IS plays a crucial role of \emph{enabler} for coping with the requirements of \emph{speed} and \emph{scale} of the activities. Here we attempt to identify possible role of IS and the essence of its link to the people and their knowledge.

\subsection{``Information Systems" and ``Knowledge"}

As noted earlier, ``knowledge" is different from ``information". Knowledge, defined in Oxford dictionary as ``the sum of what is known", is possessed by individual human beings. It is a \emph{tacit} entity, consisted of our experience, skills, expertise, mental belief and many more components. To perform a work one needs to exercise his/her cognitive faculties, which involves application of the knowledge possessed by the individual. Performing the work, in turn, adds new knowledge and/or update the knowledge of the worker. One can convey some portion of knowledge in his/her possession  by act of \textbf{Articulation} (which itself is a complicated exercise of cognitive abilities). The \emph{articulated} or \emph{potentially articulable} portion of knowledge is often called the \emph{explicit knowledge}. The explicit knowledge, as Polanyi\cite{Polanyi} famously surmised ``we know more than we can tell", is usually a small portion of the full knowledge possessed by the knower.

Sharing or communication of ``knowledge" among people is an indirect process. The possessor of knowledge articulates some parts of his/her knowledge, potentially useful to others. The ``articulated knowledge" is ``information"\cite{alavi}. A recipient can receive the information and process (again using cognitive abilities) it to gain new knowledge, depending on his/her state of existing knowledge. Gain of knowledge by the recipient is manifested in his/her \emph{improved capability} of performing works, where the knowledge is required. Thus, in true sense, \emph{information}, rather than \emph{knowledge}, is the entity amenable to sharing or communication. \textbf{Recording} of the information (i.e., articulated knowledge) in \emph{persistent media} facilitate sharing of information beyond the boundary of spatial and temporal locality of the act of articulation.

\emph{Information Systems}, as the name suggests, capable of dealing with information, actually the information (subset of the larger gamut) which are (or can be) recorded and encoded using some digital format. We must remember that \emph{the information systems have nothing to do with knowledge by themselves, unless some human possessor of knowledge uses them in their quest for information}. Even so-called \emph{knowledge-based} or \emph{intelligent} systems, when in operation on their own, work with information and data only. For accomplishing any knowledge work of even moderate complexity, it is imperative that human workers with requisite expertise participate actively and substantially in the process. Such tasks typically fall in the categories of \emph{semi-structured} and \emph{unstructured} decision making tasks\cite{Keen} and not amenable to automation. Therefore, providing effective IS support for such tasks turns essentially into creating an environment where IS and knowledge worker can complement each others' capability in a symbiotic relationship, as envisaged by Licklider\cite{Licklider} back in 1960s.

\section{Scope of IS support}

The knowledge workers use information systems to deal with \emph{information}, i.e., for \emph{accessing} information and \emph{manipulating} information for easy processing and consumption. Let us try to analyze the requirements with respect to the former. To access information, firstly they have to exist in accessible form. Some of the available information may exist only in peoples' minds, where accessing them requires interacting with the possessors. For the time being, we let us focus on the part of available information which is \emph{archived} and \emph{accessible by computer-based information systems}. Usefulness to the knowledge worker, of information retrieved from archive(s), crucially depends on three attributes of the retrieved information. They are (1) relevance, (2) linkage and (3) provenance. We discuss each of them below.

\subsection{Relevance}
The information requirement of a knowledge worker depends on various type-specific and instance-specific characteristics of the current knowledge work instance or task instance (TI). Further, the knowledge-intensive tasks are never \emph{monolithic} or \emph{atomic} in nature. They are consisted of a set of interrelated \emph{activities}, often fairly diverse in nature. At any given time, the worker is typically engaged in performing an activity as part of a larger KW instance and his/her information requirement, both content-wise as well as pattern-wise, is influenced heavily by the nature of the \emph{current} activity. However, the \emph{documents} containing the information, whether distributed across various systems or archived and managed with state-of-the-art Content Management Systems (CMS), usually have information relevant to the current activity embedded in small portions of each of a set of documents. With the advanced search technologies of recent time, we can identify fairly precisely the set of documents. But it is still left to the human worker to peruse each (or as many as humanly possible!) document, find out the relevant piece of information, understand, interpret and synthesize it with other pieces of information. Without doubt this process of information gathering and processing makes a \emph{huge demand} on the time and cognitive ability of the worker.

The challenge, the IS researchers should address in this front, is to present to the worker, based on the query and activity context, not \emph{a set of whole documents} within which useful pieces of information are distributed, but \emph{relevant information extracted from the containing documents}. As we shall see later there are several conceptual as well as technological issues to be addressed to realize this capability.

\subsection{Linkage}
The value or usefulness of a piece of information increases manyfold when it can be associated other related pieces of information and studied together. Let us consider an example: A worker is tasked to prepare a plan of a marketing campaign of a set of products. While preparing the plan, at one stage she is trying to work out the feasibility of offering some incentive to customers for boosting the sale of some products. The incentive can be of several kind, e.g., a discount scheme, an easy installment scheme etc. While studying the problem, she accesses archived information on other discount schemes offered by the company as part of earlier marketing initiatives of similar nature. However, the information that \emph{during that campaign so and so products were sold at x\% discount} is of little value to her. She would like to know the other pieces of related information, such as \emph{what were the sales and profit targets from the discount sales, whether the implementations of the schemes met the targets, if not what were the reasons, is there some interrelations among sales of different products} and many other pieces of related information.

An effective information system must offer the ability to the user the to find and navigate the links between related pieces of information. The system, along with the information, must be able to provide the \emph{linkage} among the pieces of information based on various types of relationships. One of the technological challenges towards providing the linkage is one of information integration. The state-of-the-art in this area is reviewed by Bernstein and Haas\cite{Bernstein}. Though they are able to identify several potentially important technologies, these technologies are usually applied in specialized applications. To achieve the level of integration support required for providing effective linkage information, we need to build a systematic approach to guide us on the application of the individual technologies.

\subsection{Provenance}
Provenance of a piece of information encompasses related information on \emph{how, where, what, when, why, which} and \emph{by whom} the information has come into being. It form the basis of \emph{validation} of the information. Its importance in the domains like scientific research\cite{Sahoo}, law enforcement etc. are well-established. However, in other domains also increasing importance of provenance can be related to the various compliance requirements. Apart from such legal/regulatory requirements, easy access to provenance can be of great value to the knowledge workers. Ability to check the provenance of information assures a worker of the quality of the information. Further, in scenario of collaborative works, the information, containing some crucial arguments, prepared by a co-worker can be examined much rigorously using the provenance and linkage facilities to go beyond the upfront information, into other pieces of information those form the premises of the arguments. Thus the contributions of the workers will be much more effective in their exchange of information and collaboration.

Incorporating facilities for supporting provenance in the information system is fairly complex task. It requires ability of the system to be aware of other systems, including databases, data analysis systems, ERP, Internet etc. in the organizational environment, as well as ability to communicate to many of them. Further, identity of the workers, working on various systems and responsible for development of a piece of information, also need to be associated with the information itself.

\section{The \emph{Information Warehouse}}
To the best of the authors's knowledge there are no published work focusing on the above aspects of information access together. To address the above issues together we need to look at the problem of information archival and usage from a different perspective. In the following we attempt to analyze the problem and propose a new systematic approach to reorganization and archival of the information originally available in textual documents. Let us call the proposed approach and its potential implementations, borrowing in spirit from Data Warehouse,  the \textbf{Information Warehouse} or \textbf{IW}.

We are mainly interested in the information contained in textual documents which are typically produced as the end products (strategy, plan, policy, manual etc.) as well as intermediate products (reports, notes, memos, minutes of meetings, consultation papers etc.). These documents are prepared by human workers for consumption of other human workers. Typically, each of them strive for completeness, so that the content gives a coherent picture of the issue under consideration. They refer to other sources of information which are used as premises for the arguments made in the document. Overall, the content of a document can be seen as an \emph{intellectual aggregation} of many pieces of information and result of \emph{articulation} of knowledge by the author of the document. Let us identify the organization of information in documents as the \textbf{Articulation View of Information (AVI)}.

\begin{figure*}
\epsfxsize=0.8\hsize \ArtWork{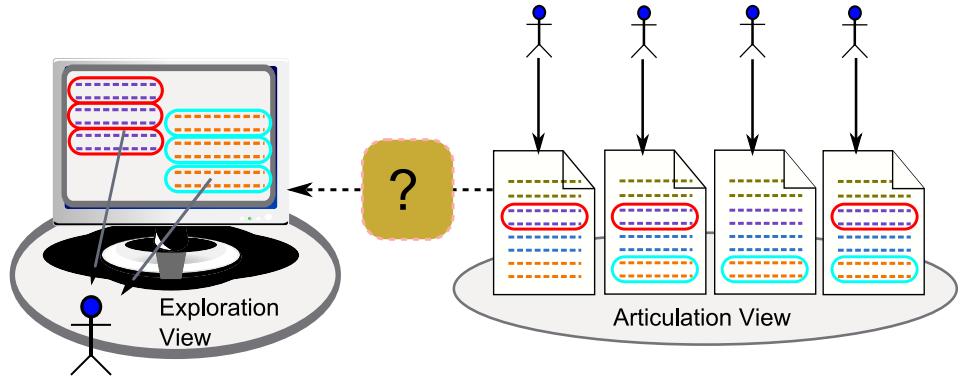} \caption{The gap between Articulation and Exploration views of information.}\label{fig:gap}
\end{figure*}

In light of the requirements laid above, it can be easily seen that the AVI is hardly optimal in satisfying the need of the workers as outlined in previous section. Let us call the view of information capable of satisfying the users' requirements the \textbf{Exploration View of Information (EVI)}. It is evident that there are significant differences between these two views. So, the problem is essentially one of \emph{computing the EVI, in response to user requirements, from the information originally available in AVI}. Figure \ref{fig:gap} depicts the situation. However, creating required EVI on-the-fly from AVI is, if not impossible, extremely difficult. One feasible solution is to build an archive, the \emph{Information Warehouse}, where the original information is re-organized to form an intermediate \textbf{Bridge View of Information (BVI)}. The bridge view should be designed such a way that the various EVI can be easily computed on-demand from it. Further, it should be easy to devise means and methods to convert information in articulation view to corresponding bridge view. Let us call this process of conversion as the \textbf{Transcription} process. In the following we expand the characteristics of the Bridge View and the Transcription process.

\subsection{\emph{Bridge View} of Information (BVI)}
\begin{figure*}
\epsfxsize=0.6\hsize \ArtWork{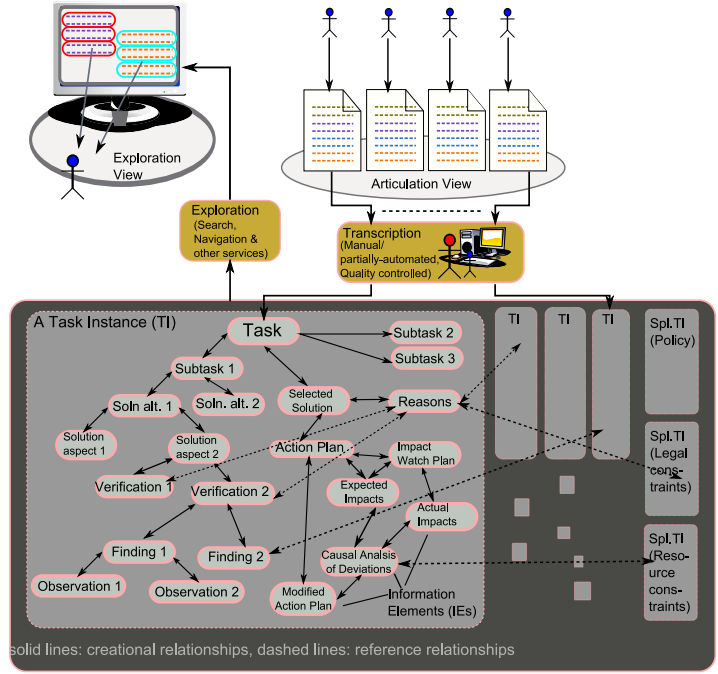} \caption{The Bridge (CaRN) view of information in form of a network of information elements.}\label{fig:bridge}
\end{figure*}

Though there is no unique approach to build a bridge view of information, here we propose a bridge view in form of a network of \emph{Information Elements (IE)}. The IEs are \emph{objects}, specialized to serve as containers of various types information developed in the course of \emph{performance of knowledge work instances}. The IEs are interconnected through two types of links, the \emph{creational} links and the \emph{reference} links, and we call the organization as the ``\textbf{C}reational \textbf{a}nd \textbf{R}eference \textbf{N}etwork" or CaRN view in short. A creational link joins two IEs when the information content of one is created to satisfy the need of creating the content of the other. On the other hand, a reference link exists between two IEs when the content of one is used, but not explicitly created to cater the need of the other. It is clear that the creational links reflect also the relationships between pieces of information developed by activities performed as part of the same KW instance. This allows us to define macro elements of the CaRN as the \emph{Task Instances} or TIs which contain all information developed during performance of a KW instance. The structure  of CaRN is depicted in Figure \ref{fig:bridge}. It can be noticed that as consequence of above definitions, the creational links are allowed to form between IEs who belong to the \emph{same} TI. On the other hand the reference links are free to cross the TI boundaries to associate IEs belonging to different TIs.

Implementations of the archival of information in CaRN view can be accomplished with relational databases, XML databases, content management systems or some hybrid of them (we are currently involved in testing several options). However, irrespective of implementation technology adopted, it can be easily perceived that in this scheme, access to the the content of IEs and their interconnections can provide the user with information with \emph{high relevance} along with capability of \emph{navigation} to other contextual information using the edges of the IE network. Further, IEs being objects, they have fields along with one containing the \emph{principal content}, other fields containing various provenance information. While creating the IEs, the provenance information can be easily stored in related fields. Thus provenance information can also be accessed by the user in the similar way as other parts of information. One more important advantage of this scheme is the possibility of using various search and analysis techniques developed in the field of \emph{graph/network analysis} to analyze the the content and structure of the archived information.

\subsection{The \emph{Transcription} process}
As pointed out earlier, the information is originally available in form of text documents (AVI) developed in the course of and as results of performance of knowledge work instances. To reorganize the information in the bridge view, the first step is to identifying sets of documents corresponding to individual KW instances. Next, the \emph{transcription process} aims to build TIs in the bridge view for the KW instances from corresponding sets  of documents. The task involves isolating pieces of information from the documents, discovering their interrelationships and map them into appropriate types of IEs and their linkage patterns. Undoubtedly this is a non-trivial task. To accomplish this we need to understand the characteristic of the knowledge work types in terms of their constituent actions, their information requirements and interdependencies. Given this understanding it is possible to build KW type-specific templates or maps and other artifacts. These can be used in several ways. In the scenario where an IW based system is already in operation, methods and processes can be established, where the workers using user-friendly interfaces, customized with help of the relevant artifacts, can easily articulate and resulting information is organized and archived in IW. The other scenario emerges when we want to transfer information from \emph{already existing} documents to the IW. Here the artifacts can be used to build tools for transcription performed by the human transcribers manually or by  \emph{partially} automated \emph{Natural Language Processing} based \emph{Text Analysis} techniques. Depending on the particulars of the environment, a mix of both scenario may emerge.

Implementations of the transcription processes have several organizational/managerial aspects. One of them is when to convert the information to into bridge view. The choice can be anything between ``as soon as the original document(s) become available" and ``when the information embedded in the documents is asked for first time". Further, the quality/accuracy of the information accessed is of crucial importance for their usability. Hence, to ensure the quality of the the information in bridge view it is essential to embed proper controls and practices for oversight within the process.

\section{Information Systems based on IW}
The \emph{information warehouse} raises possibilities of developing many exciting IT applications, including those addressing various aspects of complex \emph{information integration} problems. However, here we concentrate on the bigger picture and outline the idea of a new class of information systems, the \textbf{Knowledge work Support Systems} or \textbf{KwSS}, based on IW. As the name suggests, the systems' design can facilitate the workers on the basis of KW types (e.g., decision-making, planning, research, policy-making etc.) to be performed  in their respective domains (e.g., finance, retail, healthcare etc.). This approach offers us the scope to modularize the system design process by concentrating on each of the KW types to be supported separately. This involves including KW-specific information, such as various activities and their interrelationships in context of the respective KW types, in the system. We can also include the ``domain vocabulary" in form of \emph{thesauri} or \emph{ontology}. Use of these information enable us to annotate the archived information and present the retrieved information in easily understandable way to the end-user. Further, we can use these information to provide means for building powerful semantic-based search and processing capabilities of the archived information.

The KwSS approach of building information systems can facilitate solution of many crucial problems currently faced by organizations. For example, such systems can  provide opportunity of building efficient \emph{expertise locator}\cite{Ehrlich} applications within large organizations. The information in the IW can be used by the application to analyze the work-history/performance of the workers and build their accurate \emph{expertise profiles}. There are numerous other possibilities. The core strengths of these systems stem from two sources. Firstly, the proposed scheme of organization of information in IW allows us to archive and retrieve information with high relevance, contextual linkage and verifiable provenance. Secondly, the KwSS provides a framework of systematically (and incrementally, if required) implementing the IW and information delivery infrastructure, where a higher level of organization of archived information along with related explanatory and semantic information is realized.

To highlight the above points, here we discuss another possibility of far-reaching impact. Organizations frame their guiding principles in forms of artifacts like policies, best practices, manuals etc., which form the basis of how the businesses are conducted. Preparation of such guiding artifacts are extremely knowledge-intensive works and depend on the expertise and information available at the time of their formulation. However, in today's \emph{dynamic} world, the \emph{validity} of the information and their interpretations, which form premises of those guiding principles do not remain same over the time. Thus, there is a crucial need to periodically examine them and updating them to suit the current environment. Without doubt, this is an extremely difficult task, often associated with the \emph{double-loop learning}\cite{Argyris} capabilities of the organizations. In the KwSS approach, preparation of these important artifacts are treated as KW types and sophisticated support for performing them can be built into the system. This will allow the workers to easily access the archived information of earlier instances of similar tasks available in IW and analyze in light of the new information.

\section{Conclusion}
In this paper we have attempted to examine the problem of information access and usage by knowledge workers in organizational setups from a different perspective and provide a outline of information systems capable of addressing several issued currently not addressed adequately. We proposed the scheme of building Information Warehouse, embodying a new approach for archival and accessing information originally available in textual documents. The capabilities of IWs can form the basis of various integrated information management platforms. Further, we have proposed the KwSS framework of building IW based systems, which can exploit full power of the IWs as well as extend it into developing comprehensive support for myriad types of knowledge works across domains. They also give rise to many exciting possibilities of building systems with deep impacts in the organizational scenario. Currently I am working with a team researchers and engineers in implementing and enhancing a IW and KwSS prototype.

\end{document}